\documentclass[twocolumn,prb,showpacs,10pt,superscriptaddress]{revtex4-1}
\usepackage{textcomp}
\usepackage{amsmath}
\usepackage{amssymb}
\usepackage{subfig}

\usepackage{color}
\usepackage{multirow}
\usepackage{float} %\floatstyle{boxed} \restylefloat{figure}
\usepackage{fontenc}
\usepackage{graphicx}
\usepackage{graphics}
\usepackage[justification=centering]{caption}
\usepackage{lipsum}
\makeatletter

\newcommand{\Rmnum}[1]{\expandafter\@slowromancap\romannumeral #1@}

\begin{document}

\title{A Density Functional Theory Based Electron Transport Study of Coherent Tunneling Through Cyclic Molecules Containing  Ru and Os as Redox Active Centers}
\author{Xin Zhao}
\affiliation{Institute for Theoretical Physics, TU Wien - Vienna University of Technology, Wiedner Hauptstrasse 8-10, A-1040 Vienna, Austria}
\author{Robert Stadler}
\email[Email:]{robert.stadler@tuwien.ac.at}
\affiliation{Institute for Theoretical Physics, TU Wien - Vienna University of Technology, Wiedner Hauptstrasse 8-10, A-1040 Vienna, Austria}

\date{\today}

\begin{abstract}

In our theoretical study in which we combine a nonequilibrium Green's function (NEGF) approach with density functional theory (DFT) we investigate branched compounds containing Ru or Os metal complexes in two branches, which due to their identical or different metal centers are symmetric or asymmetric. 
In these compounds the metal atoms are connected to pyridyl anchor groups via acetylenic and phenyl spacer groups in a meta-connection.
We find there is no destructive quantum interference (DQI) feature in the transmission function near the Fermi level for the investigated molecules regardless of their symmetry, neither in their neutral states nor in their charged states. 
We map the structural characteristics of the range of molecules onto a simplified tight-binding model in order to identify the main differences between the molecules in this study and previously investigated ferrocene compounds in order to clarify the structural sources for DQI, which we found for the latter but not for the former. 
We also find that local charging on one of the branches only changes the conductance by about one order of magnitude which we explain in terms of the spatial distributions and charge-induced energy shifts of the relevant molecular orbitals for the branched compounds.
\end{abstract}

\maketitle
\subsection{Introduction}\label{sec:intro}

Under the restrictions of low bias current flowing through small molecules absorbed to metal electrodes in ultrahigh vacuum at very low temperature, the field of
single-molecule electronics ~\cite{ratner, loertscher} has become accessible for a nonequilibrium Green's function (NEGF) approach combined with density functional theory (DFT)~\cite{keldysh,atk,xue,sanvito,kristian}, which allows for an atomic interpretation of experimental results in a mechanical break-junction (MCBJ) or scanning tunneling microscope (STM) setup ~\cite{C. Joachim 1995, M. A. Reed 1997,J. Reichert 2002, R. H. M. Smit 2002}. 
Destructive quantum interference (DQI) effects ~\cite{mayor,lambert1} allow for the design of logical gates ~\cite{graphical1} and memory cells ~\cite{memory} in single molecule electronics as well as the implementation of thermoelectric devices ~\cite{fano,lambert2}, since DQI is supposed to significantly reduce the conductance in conjugated $\pi$ systems where such effects were even observed at room temperature ~\cite{molen}.

\begin{figure}
\begin{center}
\includegraphics[width=\linewidth,angle=0]{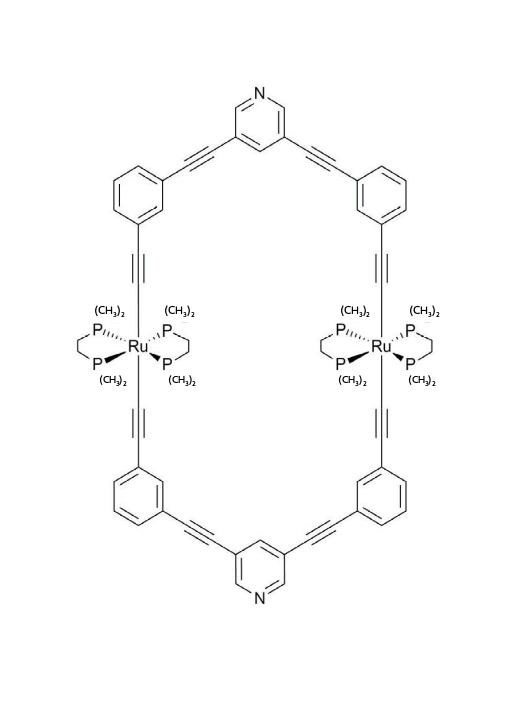}
\caption{\small Chemical structure of a cyclic Ru$_{2}$(P(CH$_{3}$)$_{2}$)$_{8}$(C$_{2}$H$_{4}$)$_{4}$(C$_{6}$H$_{4}$)$_{4}$ bis(pyridyl-diacetylide) molecule.
  \label{Fig.chemicalstru}} 
\end{center}
\end{figure}

Experimentally, the design and synthesis of branched compounds containing ferrocene moieties in each branch has been proposed ~\cite{tim} for the purpose of creating single molecule junctions, where the combination of quantum interference effects with redox gating for coherent electron tunneling as well as the electrostatic correlation between spatially distinct redox centers for electron hopping ~\cite{Georg 2015} can be explored. A detailed theoretical analysis of branched compounds containing ferrocene centers has been published in our previous work ~\cite{Xin2017}. In order to assess the generality of this analysis we apply the same models and methods to cyclic Me$_{2}$(P(CH$_{3}$)$_{2}$)$_{8}$(C$_{2}$H$_{4}$)$_{4}$(C$_{6}$H$_{4}$)$_{4}$ bis(pyridyl-diacetylide) molecules (Fig. ~\ref{Fig.chemicalstru}), where the two metal atoms Me are Ru or Os and we use the notation of Ru/Os, Os/Os, Ru/Ru for complexes containing symmetrical and asymmetrical branches. 

There are experimental and theoretical studies on coherent tunneling and electron hopping through single-branched molecules containing Ru atoms as redox-active centers ~\cite{Georg 2015, Schwarz 2015}, where Ref. ~\cite{Georg 2015} focuses on the comparison between a coherent tunneling and a hopping mechanism in dependence on the molecular length for ruthenium bis(pyridylacetylide) wires, and the work in Ref.~\cite{Schwarz 2015} investigated redox-switches with coherent tunneling for the electron transport through the junction and the switching induced by hopping. 

For the work on branched ferrocene compounds we used a simple but effective model, i.e. a  combined atomic orbital (AO)/fragment orbital (FO)-tight-binding (TB) model where we keep only the $p_{z}$ AOs for each atom in both anchoring groups and one relevant bridge FO. For the molecules in Ref. ~\cite{Xin2017} we found that the through-space coupling between the two anchor groups is the decisive parameter causing DQI effects in the interesting energy region, i.e. in the gap between the highest occupied molecular orbital (HOMO) and the lowest unoccupied molecular orbital (LUMO) and close to the Fermi level E$_{F}$.
For the molecules which are the focus of this study (Fig. ~\ref{Fig.chemicalstru}) asymmetry can be introduced by exchanging one of the  Ru redox centers with Os, and due to the longer organic bridges, the redox centers are 
more decoupled from the pyridyl anchors for these molecules than for the ferrocene compounds we studied in ~\cite{Xin2017}. In this article, we want to address two issues: i) How do the TB models we used for the branched ferrocene molecules in Ref.~\cite{Xin2017} interpret coherent electron transport for this new type of molecules? ii) Will the decreased through-space couplings due to the longer molecular length have an impact on the occurrence of DQI? 

The paper is organized as follows: Section ~\ref{computationaldetails} gives the computational details for all NEGF-DFT calculations. In section ~\ref{sec:results} we present transmission functions from these calculations for all junction geometries covered in this study and discuss their characteristics features, where comparisons of single and double branched molecules are made. In section ~\ref{sec:sources} we derive topological tight-binding models from the DFT calculations based on the scheme in our previous study where we found DQI for Ferrocene containing compounds 
and identify the structural sources for the absence of DQI in all Ru, Os containing compounds in the current study. In section ~\ref{Effect of charging} we compare the conductance from NEGF-DFT calculations for charged systems with the corresponding neutral ones for assessing the usefulness of these double-branched systems as molecular switches. We conclude with a summary in section ~\ref{sec:summary}.

\section{Computational details for the NEGF-DFT calculations}\label{computationaldetails}

For the computation of transmission probabilities $\mathcal{T}(E)$, we performed DFT calculations with a PBE~\cite{pbe} XC-functional within a NEGF framework ~\cite{atk,xue,sanvito,kristian} using a linear combination of atomic orbitals (LCAO) ~\cite{lcao} as basis set on a double zeta level with polarization functions (DZP) using the GPAW code ~\cite{J.Mortensen 2005, J.Enkovaara 2010}, where a grid spacing of 0.2 \AA{} for the sampling of the potential in the Hamiltonian on a real space grid is used.
In our transport calculations, the scattering region is formed by the respective metal organic compounds and three and four layers for the upper and lower fcc gold electrodes, respectively, in a (111) orientation and with 6 $\times$ 10 gold atoms in the unit cell within the surface plane. The distance between the Au ad-atom attached to the electrodes surface and the N atom of the pyridyl anchor groups is 2.12 \AA{} ~\cite{pyridil1} and for the $k$ points only the $\Gamma$ point is used in the scattering region for evaluating $\mathcal{T}(E)$ due to the rather large cell sizes in our simulations. 

\section{Results and discussion of the NEGF-DFT calculations}\label{sec:results}

\begin{figure}
\begin{center}
 \includegraphics[width=\linewidth,angle=0]{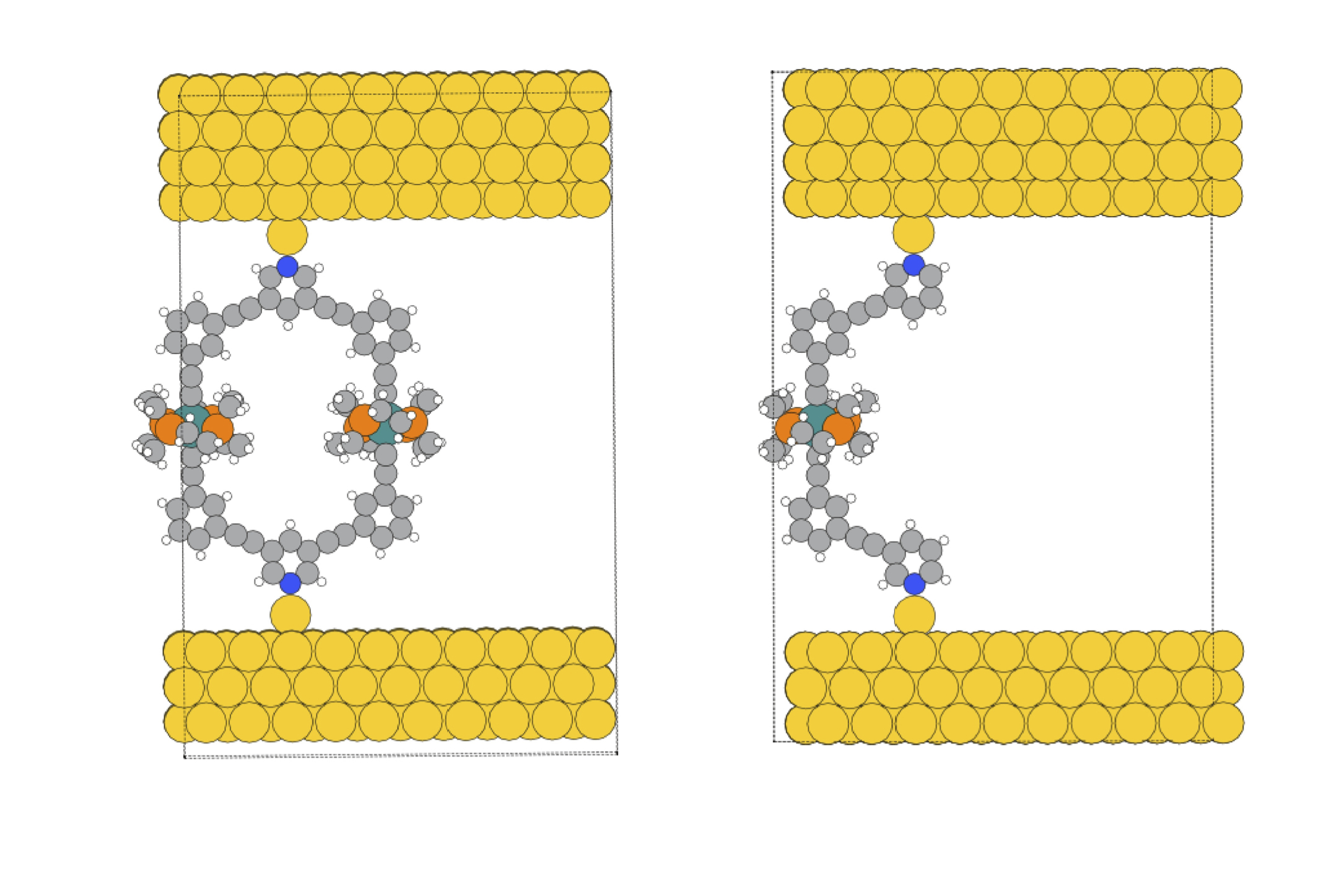}
 \caption{\small Junction setups containing the molecule in Fig. ~\ref{Fig.chemicalstru}, where for the double-branched molecule (left panel) the two metal atoms are varied as Ru/Ru, Ru/Os and Os/Os and for the single branched molecule (right panel) the single metal atom is either Ru or Os.}
 \label{Fig.neutral_big_cyclicmol_junctions} 
\end{center}
\end{figure}

\begin{figure}
\begin{center}
\includegraphics[width=1.0\linewidth,angle=0]{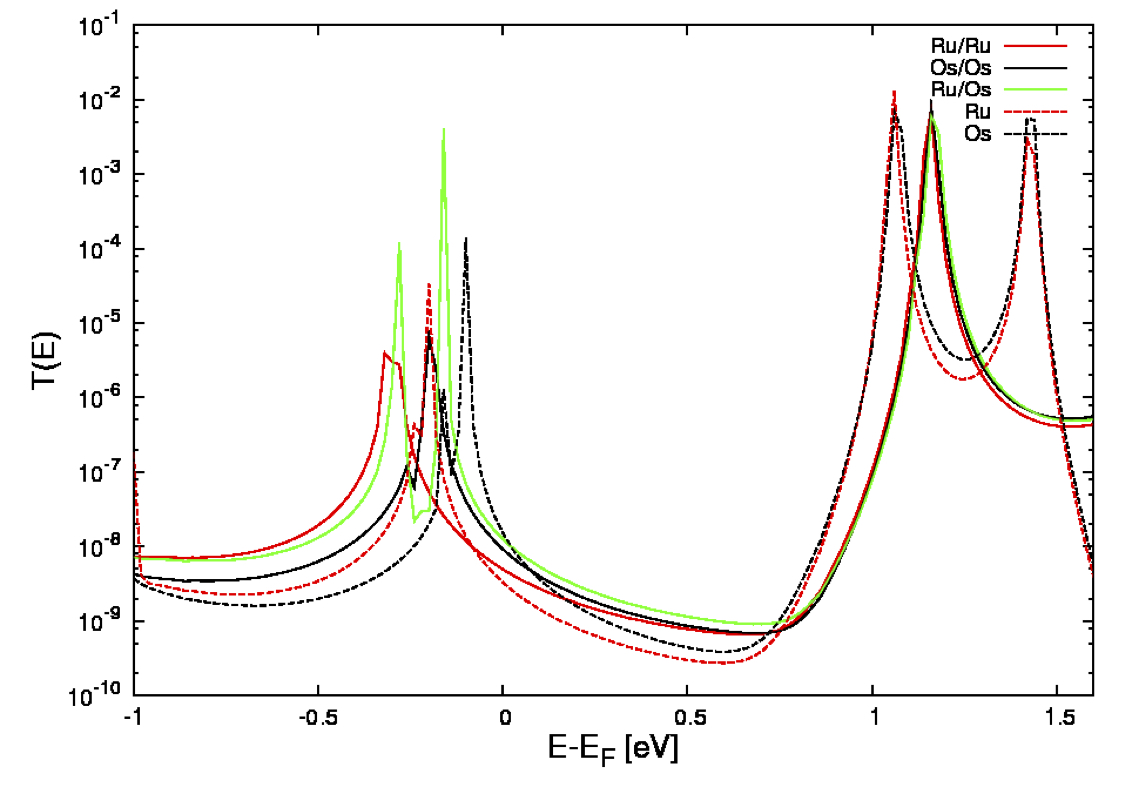}
 \caption{\small Transmission functions from NEGF-DFT calculations for the double-branched molecules Ru/Ru (solid red line), Os/Os (solid black line), Ru/Os (green solid line) as well as for single-branched Ru (dashed red line) and Os (dashed black line) in the junction setup shown in Fig. ~\ref{Fig.neutral_big_cyclicmol_junctions} in their neutral states, respectively.}
  \label{Fig.big_cyclicmol_tf} 
\end{center}
\end{figure}

In Fig. ~\ref{Fig.neutral_big_cyclicmol_junctions} we illustrate the molecular junctions derived from the compound in Fig. ~\ref{Fig.chemicalstru}, 
where we vary the two metal atoms acting as redox-centers to be Ru/Os, Os/Os and Ru/Ru.
In the resulting transmission functions $\mathcal{T}(E)$ for all three combinations and for comparison also compounds with single-branched cases of Ru and Os, which we show in Fig.~\ref{Fig.big_cyclicmol_tf}, the HOMO peaks are close to the Fermi level for all investigated systems. Hence we expect the conductance to be dominated by the MOs below $E_{F}$. 

Our definition of DQI is that the transmission through a system with more than one MO around E$_F$ is lower than the sum of the individual contributions of these MOs to $\mathcal{T}(E)$ ~\cite{pyridil3, victor}. 
The exact energetic position of the Fermi energy within the HOMO-LUMO gap, which is also affected by the underestimation of this gap in our DFT calculations with a semi-local parametrization of the XC functional, will have a crucial impact on the quantitative value obtained for the conductance but qualitatively DQI will always result in a significant conductance lowering for the structures where it occurs regardless of the details of the Fermi level alignment ~\cite{victor}.

\begin{figure}
\begin{center}
\includegraphics[width=1.0\linewidth,angle=0]{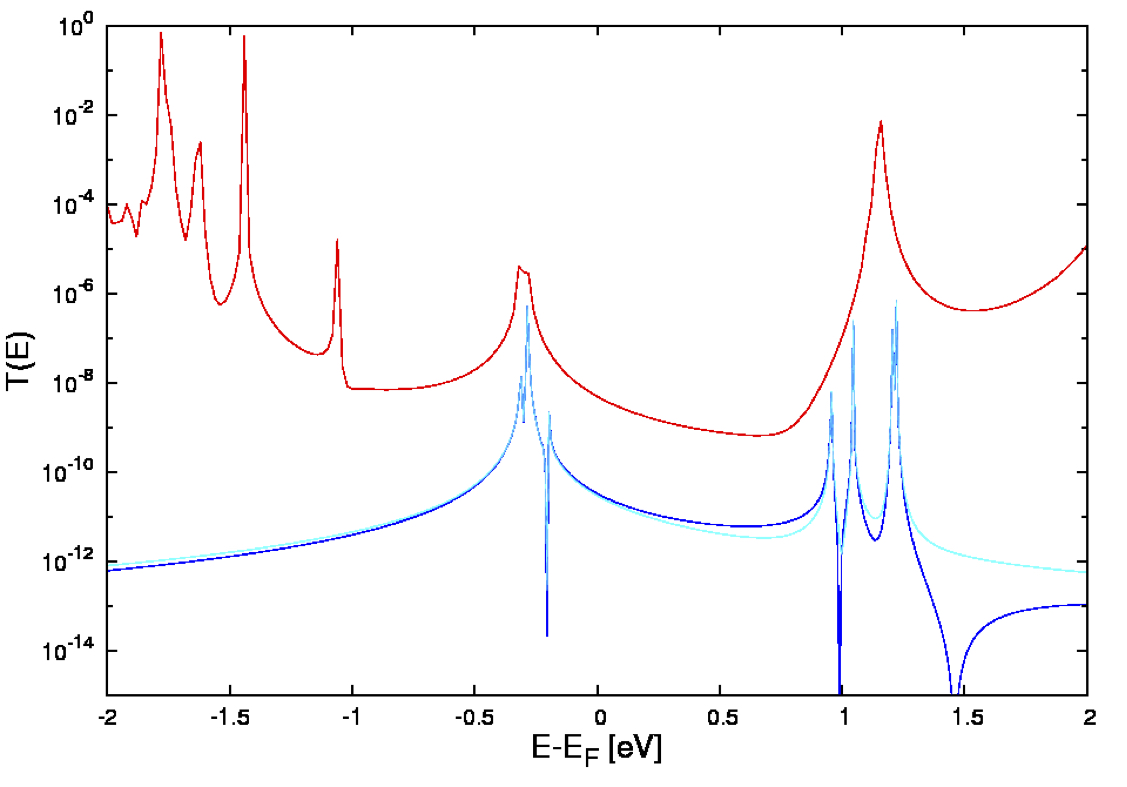}
\caption{\small Transmission functions calculated from NEGF-DFT (red curve) and Larsson's formula for the system Ru/Ru, where for the latter the blue curve denotes (HOMO + LUMO)$^2$ and the cyan curve HOMO$^2$ + LUMO$^2$.}
\label{Fig.proof_noDQI}
\end{center}
\end{figure}

We note that for molecule Ru/Os (green curve, Fig. ~\ref{Fig.big_cyclicmol_tf}) the peak splitting in the occupied region is distinct due to the intrinsic asymmetry caused by the different metal atoms in the two branches. In order to clarify whether there are DQI effects in the region of the HOMO-LUMO gap, we employ a simple TB model with a MO basis, where the eigenenergies of the molecular orbitals and their individual coupling values to the electrodes can be obtained by diagonalizing the subspace of the transport Hamiltonian with the basis functions centered on the molecule~\cite{Xin2017,pyridil4}. We use Larsson's formula ~\cite{lars, lars1, lars2, lars3} to calculate an approximation for the transmission function $\mathcal{T}(E)$, where only the frontier orbitals, namely the HOMO and the LUMO are included. 
As one can see from Fig. ~\ref{Fig.proof_noDQI} the contributions from only  the frontier molecular orbitals reproduce the characteristic features (blue curve) of the DFT result (red curve) for molecule Ru/Ru.
From Fig. ~\ref{Fig.proof_noDQI} we also see that when the HOMO and LUMO both are coupled to the electrodes ( (HOMO + LUMO)$^2$, blue curve) we achieve identical result as from the individual contributions (HOMO$^2$ + LUMO$^2$, cyan curve) around E$_{F}$, 
which means no DQI of electron transport  through these two MOs is occurring for the system Ru/Ru. For all other systems, namely Ru/Os, Os/Os, single-branched Ru and Os, we obtain the same results (not shown here). 

We also compare the transmission functions for single (red and black dashed lines) and double-branched (red and black solid lines) molecules to illustrate the impact of the number of branches in Fig. ~\ref{Fig.big_cyclicmol_tf}. As we can see the transmission functions for the Ru (dashed red line) and Ru/Ru (solid red line) molecules are rather similar, and the zero-bias conductances differ only by a factor of about 1.5. The amount of peaks in the occupied region for the double-branched molecule is higher than the one in the single-branched molecule as there are more molecular states coupled to the electrodes for the former. Comparing Os (dashed black line) and Os/Os (solid black line), we find the same qualitative differences but the conductance for the double-branched molecule is now only higher by about a factor of 0.6 for the single-branched molecule. 

The next question is then what is the reason for the absence of DQI in the HOMO-LUMO gap when these molecules share the design ideas with the ferrocene molecules we studied before ~\cite{Xin2017}. In order to address this question we apply the AO-FO model we previously used for the ferrocene systems ~\cite{Xin2017} in the next section and take the single-branched  Ru molecule as an example due to its representative transmission function, which is very similar to those of the double-branched Ru/Ru, Os/Os, Ru/Os and single-branched Os molecules. 

\begin{figure}
\begin{center}
\includegraphics[width=1.0\linewidth,angle=0]{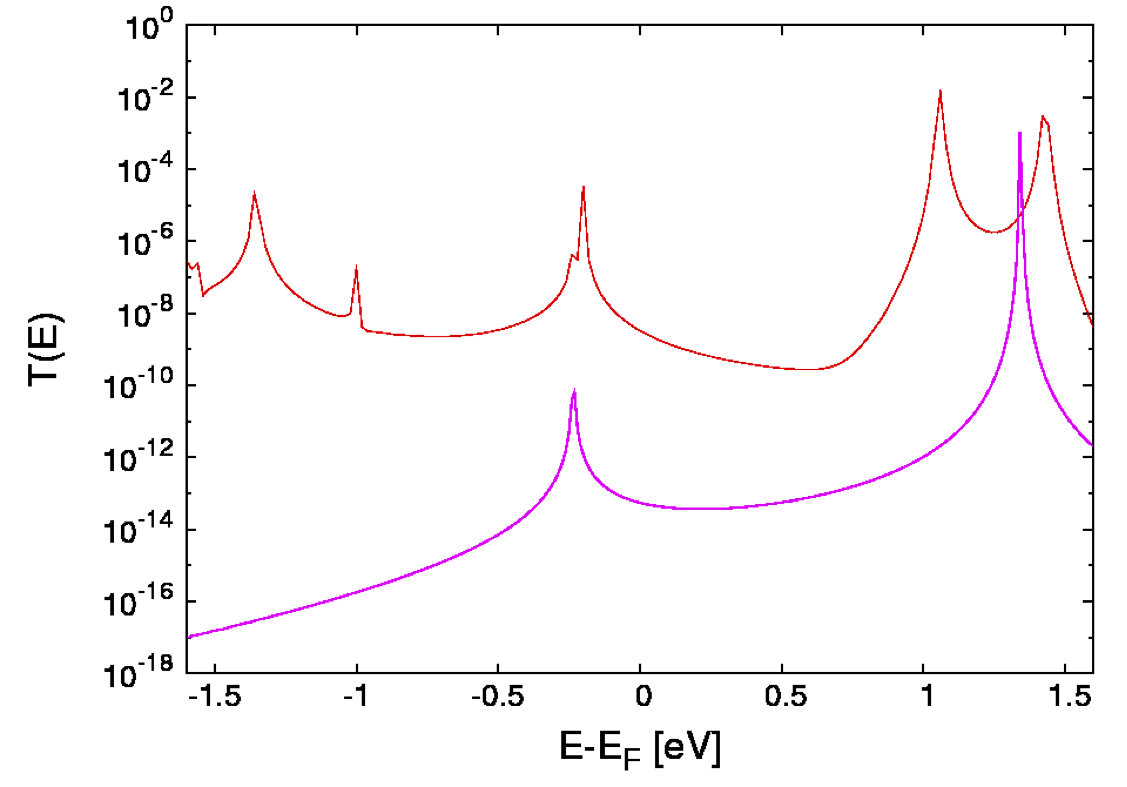}
 \caption{\small Transmission functions for a single-branched Ru molecule, where the red curve is from a NEGF-DFT calculation. The magenta curve is obtained from a NEGF-TB calculation with one FO anchor state on each side and one bridge FO, where the anchor FO is obtained from a subdiagonalization of the transport Hamiltonian containing the $p_{z}$ state of each carbon or nitrogen atom. For the electrodes a chain of single AOs with onsite energies of 0.83 eV and coupling values of -5.67 eV within the electrodes is used.}\label{Fig.single_Ru_tf_AO_FO}
\end{center}
\end{figure}

\section{Comparison of simplified 3 $\times$ 3 Hamiltonians for  Single- branched Ferrocene and Ruthenium containing molecules} \label{sec:sources}

Based on the scheme we developed in Ref. ~\cite{Xin2017} we use the simplified 3 $\times$ 3 Hamiltonian in order to obtain a mathematical perspective for explaining the role of essential structural features for the occurrence or absence of DQI according to the following procedure, where we take single-branched Ru as an example. In a first step we define the $p_{z}$ AOs of the anchor groups containing pyridyl and the attached acetylene moieties by diagonalizing the subspace of the transport Hamiltonian corresponding to each carbon or nitrogen atom on these groups and picking the $p_{z}$ states, which can be identified by their onsite energies and symmetry. For the bridge group two relevant bridge FOs (again obtained by a subdiagonalisation) in the occupied region are considered where we define Ru plus Phosphine ligands and conjugated spacers including the acetylene and benzene groups as part of the bridge. Then we diagonalize the two anchor group subspaces now defined only by p$_z$ orbitals in order to get the relevant anchor FOs on both sides. We then minimize the Hamiltonian to the most simple one, which only contains the three most relevant states, i.e. one FO on each anchor group and one bridge FO. We show $\mathcal{T}(E)$ obtained from NEGF-TB calculations for such a 3 FO model in Fig. ~\ref{Fig.single_Ru_tf_AO_FO}, and find that if qualitatively reproduces the characteristic features found in NEGF-DFT calculations.

The most distinct differences between the molecules in Fig. ~\ref{Fig.chemicalstru}  and those studied in Ref. ~\cite{Xin2017} are the molecular length and the type of metal centers. 
In our previous work ~\cite{Xin2017} we found that in order to observe a DQI feature close to E$_{F}$, the through-space coupling $t_{D}$ needs to be neither too small nor too big in size so that the DQI feature will not be pushed outside the relevant energy region around  E$_{F}$.

For a direct comparison of the molecules in our present study with those from our previous study we define the 3 $\times$ 3 Hamiltonian obtained from the 3 FO model described above as
\[
   H_{mol}=
  \left[ {\begin{array}{ccc}
   \varepsilon_{L} & t_{L} & t_{D}\\
   t_{L} & \varepsilon_{B} & t_{R}\\
   t_{D} & t_{R} & \varepsilon_{R}\\
  \end{array} } \right]
\]
for the single-branched Ru, Os and ferrocene (Fc)  systems in Table ~\ref{tab:couplings}, where $\varepsilon_{L,B,R}$ are the respective onsite energies of the three FOs, $t_{L,R}$ the electronic couplings between the two anchor FOs and the bridge FO and $t_D$ the direct coupling between the anchor FOs.

\begin {table}
\setlength{\tabcolsep}{1.1em}
\caption {Parameters entering the 3 $\times$ 3 Hamiltonian formed by three FOs ( Fig.~\ref{Fig:3FOs}) for three single-branched systems, where all values are given in eV. The values for Fc are obtained from the calculations presented in Ref. ~\cite{Xin2017}.}\label{tab:couplings}
\begin{center}
    \begin{tabular}{c c c c}
    \hline
    \hline
    %&\multicolumn{1}{ c }{} &\\ 
    
      & Ru & Os & Fc   \\
    \hline
     t$_L$ & -0.026  & -0.023   &  0.27   \\
    \hline
     t$_R$ & 0.019  & 0.019 &  -0.22  \\
    \hline
    t$_D$ & 5.7E-05 & -5.1E-05  & -0.023 \\
    \hline
    $\Delta\varepsilon$ & -1.5 & -1.5  & 0.6 \\
    \hline
    \hline
    \end{tabular}
\end{center}
\end {table}

We can see that the coupling values of t$_{L}$, t$_{R}$ and t$_{D}$ connecting the three FOs for the three systems Ru, Os and Fc differ in: i) the size of the couplings t$_{L/R}$ between anchor and bridge, which are one magnitude smaller for Ru and Os compared with the ones in Fc; ii) the through-spacing coupling t$_{D}$, which are three magnitudes smaller for the Ru and Os molecules, and iii) the onsite energy $\varepsilon_{B}$ of the bridge FO we used for the 3 $\times$ 3 Hamiltonian, which is the highest-lying FO in the occupied region for Ru and Os and was the lowest-lying FO in the unoccupied region for Fc. As a result the energy difference between anchor and bridge states $\Delta\varepsilon$ are larger in size for Ru and Os and the sign differs compared with the ferrocene molecule. 

\begin{figure}
    
    \centering
    \subfloat[]{%
     \includegraphics[clip,width=0.8\columnwidth]{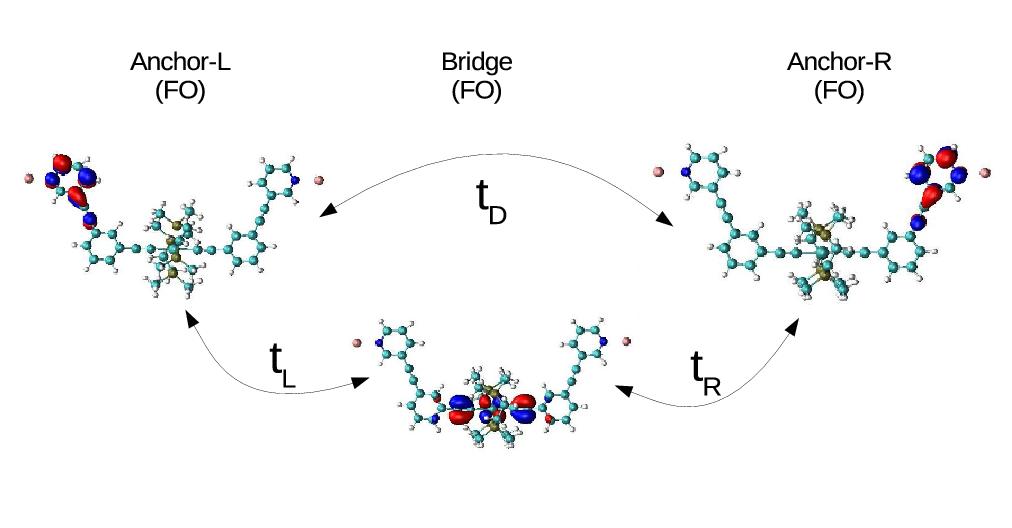}%
     }

    \subfloat[]{%
     \includegraphics[clip,width=0.8\columnwidth]{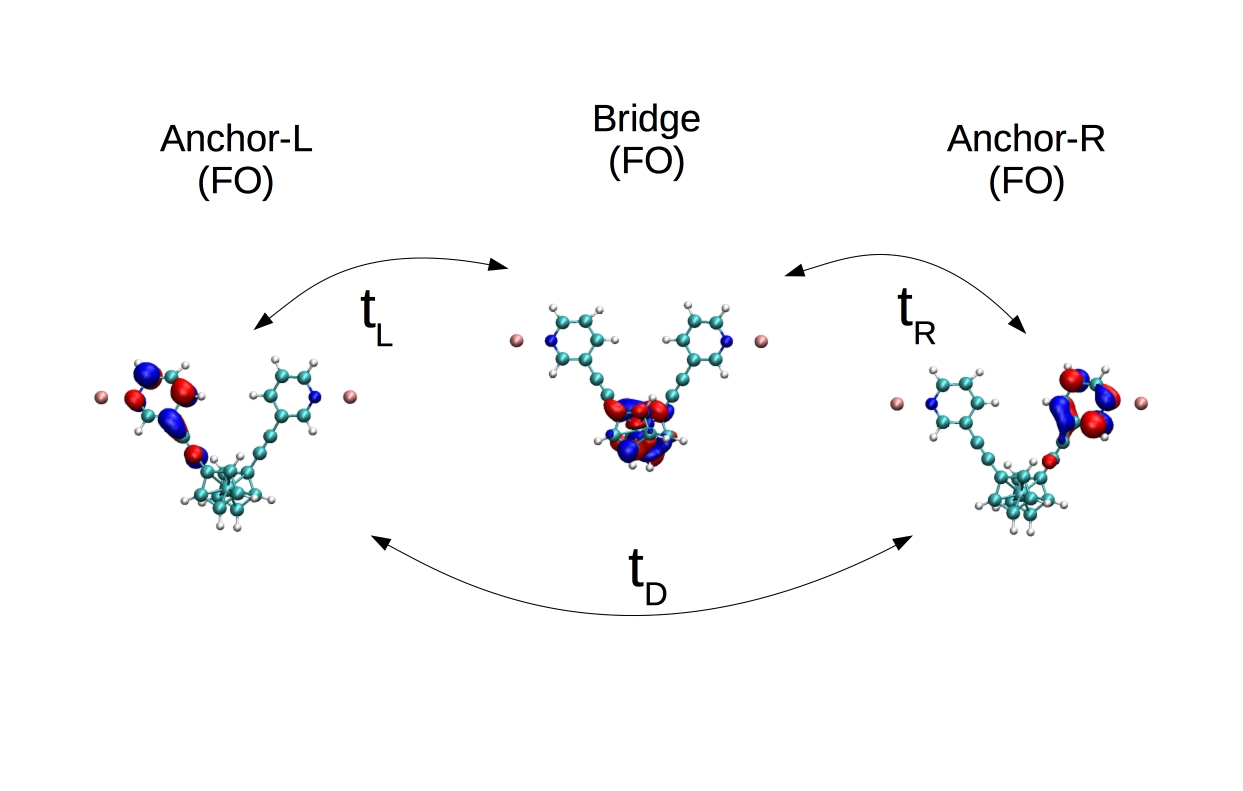}%
     }
    \captionsetup[subfigure]{}
    \caption{\small {Spatial distributions of the three FOs in the simplified model for a) single-branched Ru with the anchor FO on each side at 1.3 eV and the bridge FO at -0.2 eV; b) single-branched Fc with the anchor FO on each side at 1.05 eV and the bridge FO at 1.66 eV ~\cite{Xin2017}.}}
     \label{Fig:3FOs}
\end{figure}

The differences in couplings can be interpreted by visualizing the FOs in Fig. ~\ref{Fig:3FOs}, where the symmetry of the FOs on the anchors in Fig. ~\ref{Fig:3FOs}(a) for the Ru molecule is equivalent to what is found for the ferrocene molecule (Fig. ~\ref{Fig:3FOs} b)), but t$_{L/R}$ is decreased markedly, because in between the metal complex and the pyridyl group there is now a benzene unit separating the two for the Ru molecule. The benzene groups are defined as part of the bridge in our subdiagonalization but the resulting bridge FOs close to E$_{F}$ show no localization on them. In addition, the increased length also strongly reduces the direct coupling t$_{D}$ between the two anchor groups. The bridge state on the Ru molecule is also localized on the adjacent triple bonds, while for Fc the state is confined to the ferrocene moiety. 

Having established above that the parameters t$_{L/R}$, t$_{D}$ and $\Delta \varepsilon$ distinguish the ferrocene molecule Fc with a DQI feature close to the LUMO from the single-branched systems Ru and Os, we want to further explore the relative importance of these parameters, where we focus on the comparison of Ru and Fc regarding the three parameters entering the Hamiltonian. 
If we mark the three parameters for ferrocene molecule Fc as F$_{1}$, F$_{2}$, F$_{3}$ and for Ru as R$_{1}$, R$_{2}$, R$_{3}$, where the numbers refer to those given in Table ~\ref{tab:parameters},
there are six possible combinations of them for forming a 3 $\times$ 3 Hamiltonian. From these Hamiltonians we calculate the transmission functions with NEGF-TB for each combination in order to identify the decisive parameters enabling DQI.

\begin{figure}
    \begin{center}
    \includegraphics[width=\linewidth, angle=0]{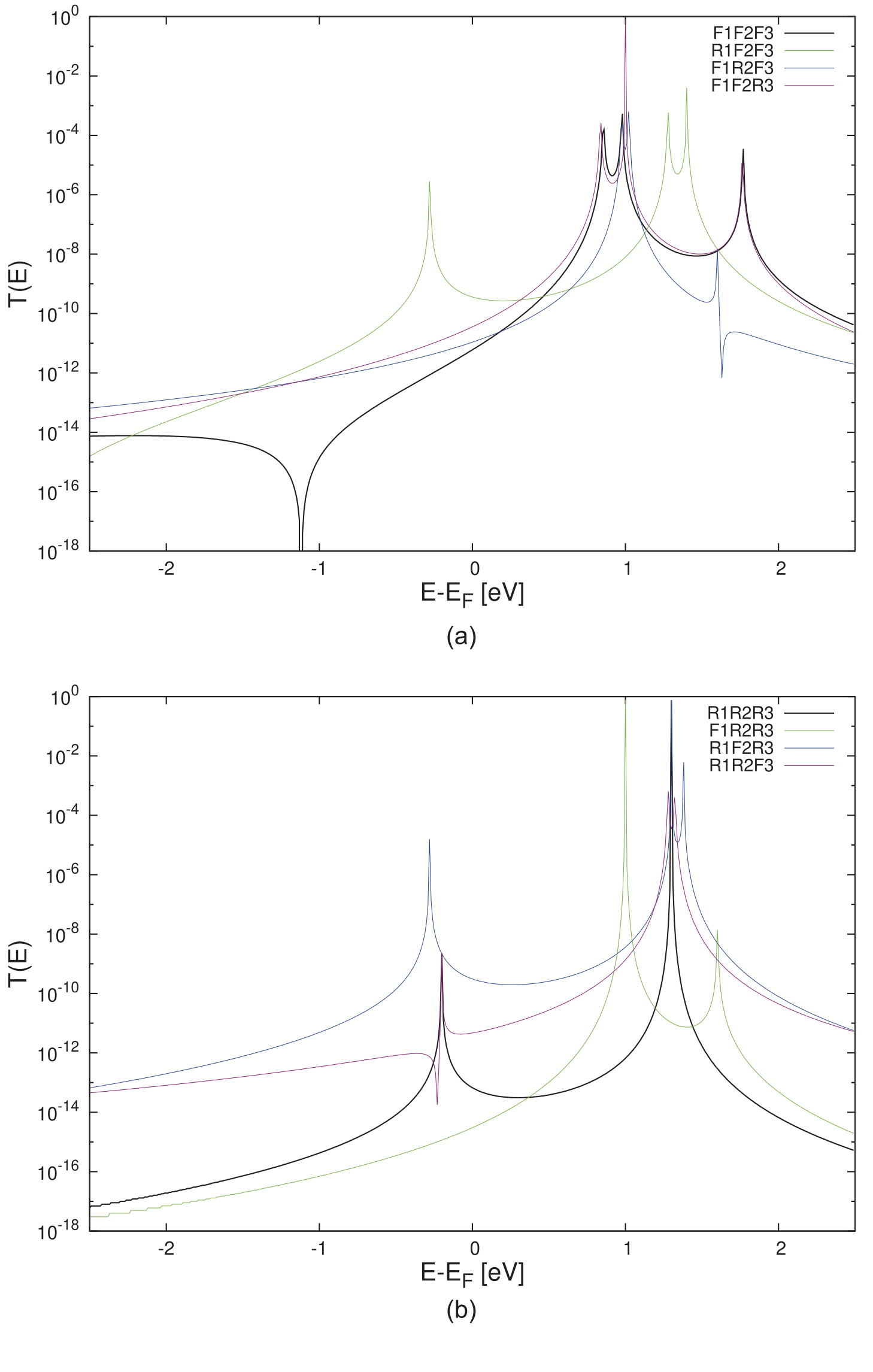}
    \caption{\small Transmission functions obtained from NEGF-TB calculations using the 3FO-model of the single-branched a) Fc molecule (black solid line) and b) Ru molecule (black solid line), where the symbols F and R represent Fc and Ru, respectively, and the indices 1-3 refer to the numbers in Table ~\ref{tab:parameters}. In panel a) we start from the parametrization of Fc in Table ~\ref{tab:parameters} (F$_1$F$_2$F$_3$) and replace each of the parameters individually by the one corresponding to Ru, and calculate $T(E)$ for R$_1$F$_2$F$_3$ (green line), F$_1$R$_2$F$_3$ (blue line) and F$_1$F$_2$R$_3$ (cyan line). In panel b) we permute the parameters in the opposite direction starting from the parametrisation of Ru in Table ~\ref{tab:parameters} (R$_1$R$_2$R$_3$) and obtaning $T(E)$ for F$_1$R$_2$R$_3$ (green line), R$_1$F$_2$R$_3$ (blue line) and R$_1$R$_2$F$_3$ (cyan line).}
    \label{tf-six-Hamiltonian}
    \end{center}
\end{figure}

For the sake of simplicity we approximate the coupling values t$_{L/R}$ with 0.025 eV and 0.25 eV for Ru and Fc, respectively, and plot the transmission functions for the six resulting Hamiltonians based on the different combinations of the three parameters in Fig. ~\ref{tf-six-Hamiltonian}.
\begin{table}
\begin{center}
\setlength{\tabcolsep}{1.1em}
\caption {The three parameters defining the 3 $\times$ 3 Hamiltonian within the 3FO-model which qualitatively reproduce the transmission function for the single-branched Ru and Fc systems, respectively.}\label{tab:parameters}
    \begin{tabular}{c c c}
    \hline
    \hline
       & Ru & Fc   \\
    \hline
     $\Delta \varepsilon$ (parameter 1) & -1.5 & 0.6 \\
    \hline
     t$_{L/R}$ (parameter 2) & 0.025  & 0.25  \\
    \hline
      t$_{D}$ (parameter 3) & 5.0E-05  &  -0.023 \\
    \hline
    t$_{D}$/t$_{L/R}$  & -2.0E-03& -0.092 \\
    \hline
    \end{tabular}
\end{center}
\end {table} 

As one can see modifying any one of the three parameters in the Hamiltonian of Fc leads to the disappearance of the DQI feature in the interesting region, which indicates that all three parameters have a fundamental influence on the DQI absence for Ru in the relevant energy region and its their interplay which is decisive for the absence or occurrence of DQI features within the HOMO-LUMO gap. The through-space coupling t$_{D}$i, however is still special in the sense that taking the value from Fc in the Ru Hamiltonian (R$_1$R$_2$F$_3$ in Fig. ~\ref{tf-six-Hamiltonian}b), induces a DQI minimum slightly below the HOMO peak.

\begin{figure}
    \begin{center}
    \includegraphics[width=\linewidth,angle=0]{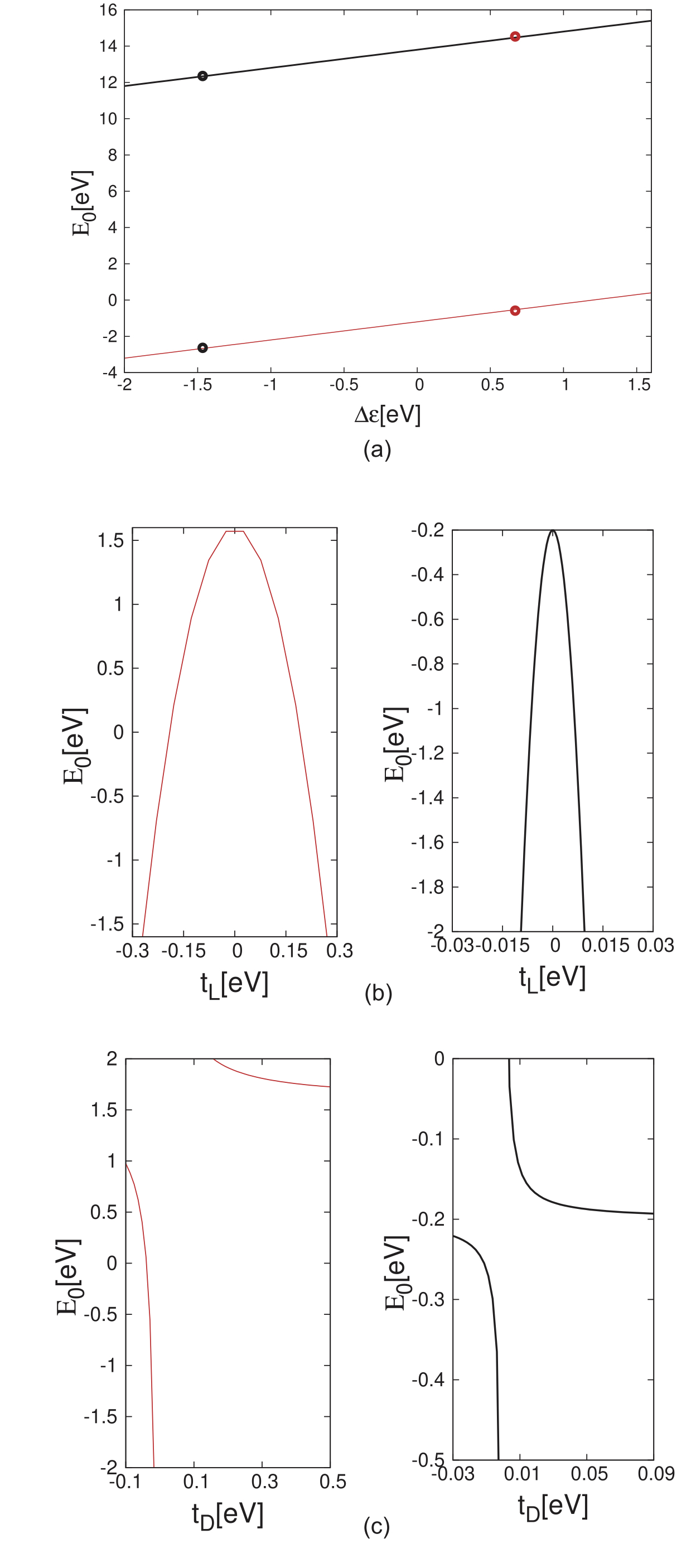}
    \caption{\small E$_{0}$ versus a) the energy difference $\Delta \varepsilon$ between the anchor and bridge states; b) the coupling value t$_{L/R}$ and c) t$_{D}$ for Ru (black curve) and Fc (red curve), respectively. In panel a) the E$_{0}$ values resulting from $\Delta \varepsilon$ (Ru) and $\Delta \varepsilon$ (Fc) at -1.5 and 0.6 eV are also marked as black and red dots, respectively.}
     \label{E0_ration}
     \end{center}
\end{figure}

In the following we keep two parameters fixed and vary one in a systematic way to further investigate  the role each parameter plays. In Fig. ~\ref{E0_ration} we illustrate the relation between each of these three parameters and the energy position of the DQI-induced minimum E$_{0}$, which we obtain from the  eigenenergies and amplitudes at the contact sites of the three MOs resulting from a diagonalization of the parametrized 3 $\times$ 3 Hamiltonian corresponding to the 3 FO-model in combination with Larsson's formula  ~\cite{victor} using the procedure described in detail in Ref. ~\cite{Xin2017}.

The such obtained single parameter dependencies can be summarized as linear for E$_{0}$($\Delta \varepsilon$) (Fig~\ref{E0_ration}a), as quadratic for E$_{0}$(t$_{L/R}$) (Fig~\ref{E0_ration}b), and as multiplicative inverse for E$_{0}$(t$_{D}$) (Fig~\ref{E0_ration}c). The black curve in Fig. ~\ref{E0_ration}a illustrates that for the single-branched Ru system with t$_{D}$ and t$_{L/R}$ fixed to the values in Table ~\ref{tab:parameters}, E$_{0}$ lies always above $\sim$ 12 eV regardless of the variation of $\Delta \varepsilon$. For the Fc system (red curve in Fig. ~\ref{E0_ration}) E$_{0}$ is around -1.1 eV, i.e. close to E$_{F}$, for $\Delta \varepsilon$ chosen as in the real system but is pushed below -3 eV when the value is replaced by the one corresponding to the Ru molecule. 

While E$_{0}$ shows a significant dependence on t$_{L/R}$ in Fig~\ref{E0_ration}b for both systems with a maximum at t$_{L/R}$=0, they differ strongly in the sense that this maximum which is defined by $\varepsilon_B$ corresponds to the LUMO peak for Fc and the HOMO peak for Ru, meaning that a variation of t$_{L/R}$ only allows to introduce a DQI minimum into the HOMO-LUMO gap for the former but not for the latter. A variation of the parameter t$_D$ on the other side allows for E$_{0}$ to cross the HOMO-LUMO gap for both systems as ca be seen from Fig~\ref{E0_ration}c albeit for values about an order of magnitude smaller for Ru when compared with Fc and for different signs.

\begin{figure}
    \begin{center}
    \includegraphics[width=\linewidth, angle=0]{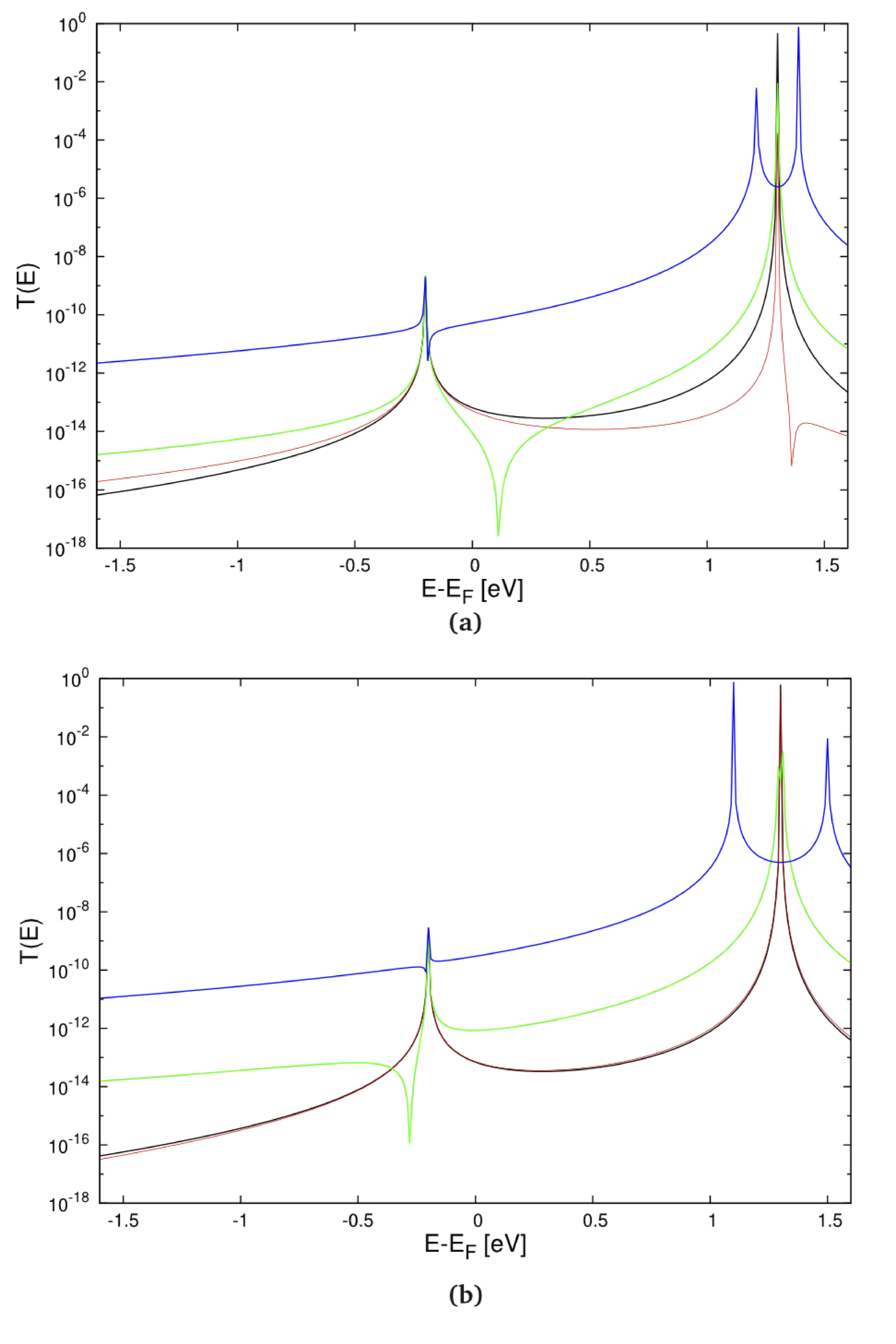}
    \caption{\small Transmission functions calculated with NEGF-TB for the single-branched Ru system for a) t$_{D}$ = 0.00005 eV (black line), 0.0004 eV (red line), 0.002 eV (green) and 0.09 eV (blue curve), and for b) t$_{D}$ = -0.00005 eV (black line), -0.0001 eV (red line), -0.008 eV (green line) and -0.2 eV (blue line), respectively.}
     \label{test_ration}
     \end{center}
\end{figure}

We put our findings on the t$_D$ dependence of E$_{0}$ to a test by performing NEGF-TB calculations for a range of values of t$_D$ where $\Delta \varepsilon$ and t$_{L/R}$ have been kept fixed to the values of the Ru Hamiltonian in Table ~\ref{tab:parameters}. In Fig.~\ref{test_ration}a we show that the real value for Ru in Table ~\ref{tab:parameters} (0.00005 eV) is too small to cause DQI anywhere near the gap, a value largeer by about one order of magnitude (0.0004 eV) makes a DQI feature appear above the LUMO and if the value is too large (0.09 eV), the feature merges with the HOMO peak. An intermediate value (0.0002 eV) places the DQI induced minimum optimally within the gap and close to the Fermi level, where its influence on the conductance will be most pronounced. In Fig.~\ref{test_ration}b we confirm the findings of Fig~\ref{E0_ration}c, namely that the sign of t$_D$ matters significantly for the location of E$_{0}$.

For a physical interpretation of our results we note that the ratio t$_{D}$/t$_{L/R}$ we obtain from the optimal t$_D$ value in Fig.~\ref{test_ration}a (0.002 eV) is 0.08 meaning that its magnitude is very close to that for the Fc system and about two orders of magnitude larger than the real value of t$_D$ for the Ru molecule in Table~\ref{tab:parameters}. It is intuitively plausible that this ratio plays such a significant role for the occurrence of DQI, where since it is a wave phenomenon destructive interference needs two different path ways which are highly asymmetrical with respect to each other but not more than one order of magnitude apart in their respective couplings. In our model these two path ways are represented by the transport through the metal center via t$_{L/R}$ and the transport from anchor to anchor mediated by t$_{D}$ as also illustrated in Fig.~\ref{Fig:3FOs}. This seems to be a general rule regardless of the detailed quantitative values for $\Delta \varepsilon$, t$_{L/R}$ and t$_{D}$, which will facilitate the chemical design of molecules enabling DQI in their electron transport characteristics in the future.

\section{Effect of Charging}\label{Effect of charging}

\begin{figure}
    \begin{center}
    \includegraphics[width=\linewidth,angle=0]{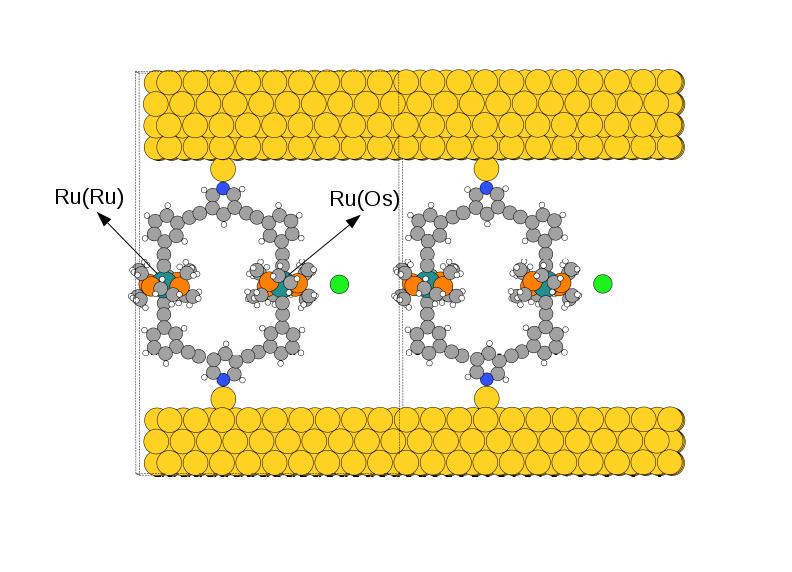}
    \caption{\small Junction geometry for two neighbouring cells in the periodic setup for the scattering region of double branched Ru/Os containing a chlorine atom for achieving local charging with a distance of 5.2 \AA{} between Cl and the closer-lying metal atom.}
    \label{charged-bigmol-junctiond-stru}
    \end{center}
\end{figure}

In order to ensure the charge neutrality in the unit cell of the system, which is necessary also for a junction with a charged molecule when applying periodic boundary conditions for electronic-structure calculations, the countercharge to the complex has to be an explicit part of the cell, where we use Cl$^{-}$ as a counterion (Fig. ~\ref{charged-bigmol-junctiond-stru}). We used the generalized $\Delta$ SCF method for calculating the charging effect in this section, where one additional electron on the chlorine $p$ shell~\cite{pyridil4} is subtracted from the molecule. In this way the molecule is charged and we keep the neutrality of the unit cell in our calculations. This approach makes use of the flexibility of  the generalized $\Delta$ SCF method to define the spatial expansion of an orbital which is forced to contain an electron as an arbitrary linear combination of Bloch states~\cite{deltascf1,deltascf2} and in our calculations is localized on a single Cl atom only as is predefined at the beginning of each iteration step. The self-consistency cycle then progresses as usual, but with the electron density of this particular orbital as a contribution to the external potential. In this way we can fix the electron occupation for the Cl manually, which solves the self-interaction problem implicitly and makes this method ideal for introducing localized charges into a junction~\cite{pyridil4}.

\begin{figure*}
    \begin{center}
    \includegraphics[width=\linewidth,angle=0]{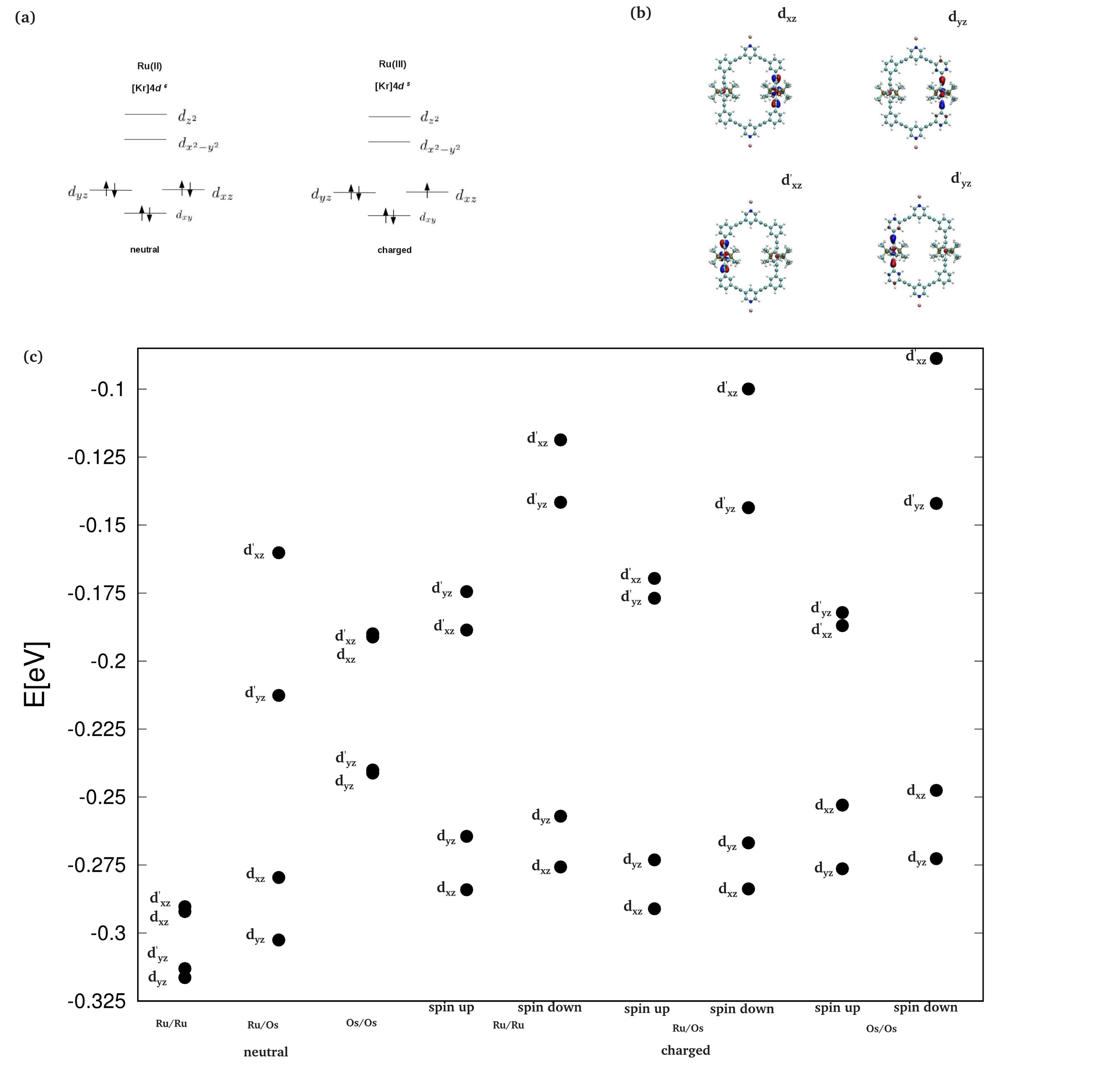}
    \caption{\small a) Electron occupation of the Ru $d$ orbitals n a low-spin configuration for neutral and charged molecules according to ligand field theory, b) spatial localization of the four relevant occupied orbitals and c) their corresponding energies within the junction and with respect to the Fermi level for three systems, namely Ru/Ru, Ru/Os and Os/Os in their neutral and charged states, respectively.}
    \label{spin_eigenenergies}
    \end{center}
\end{figure*}

\begin{figure}
    \begin{center}
    \includegraphics[width=\linewidth,angle=0]{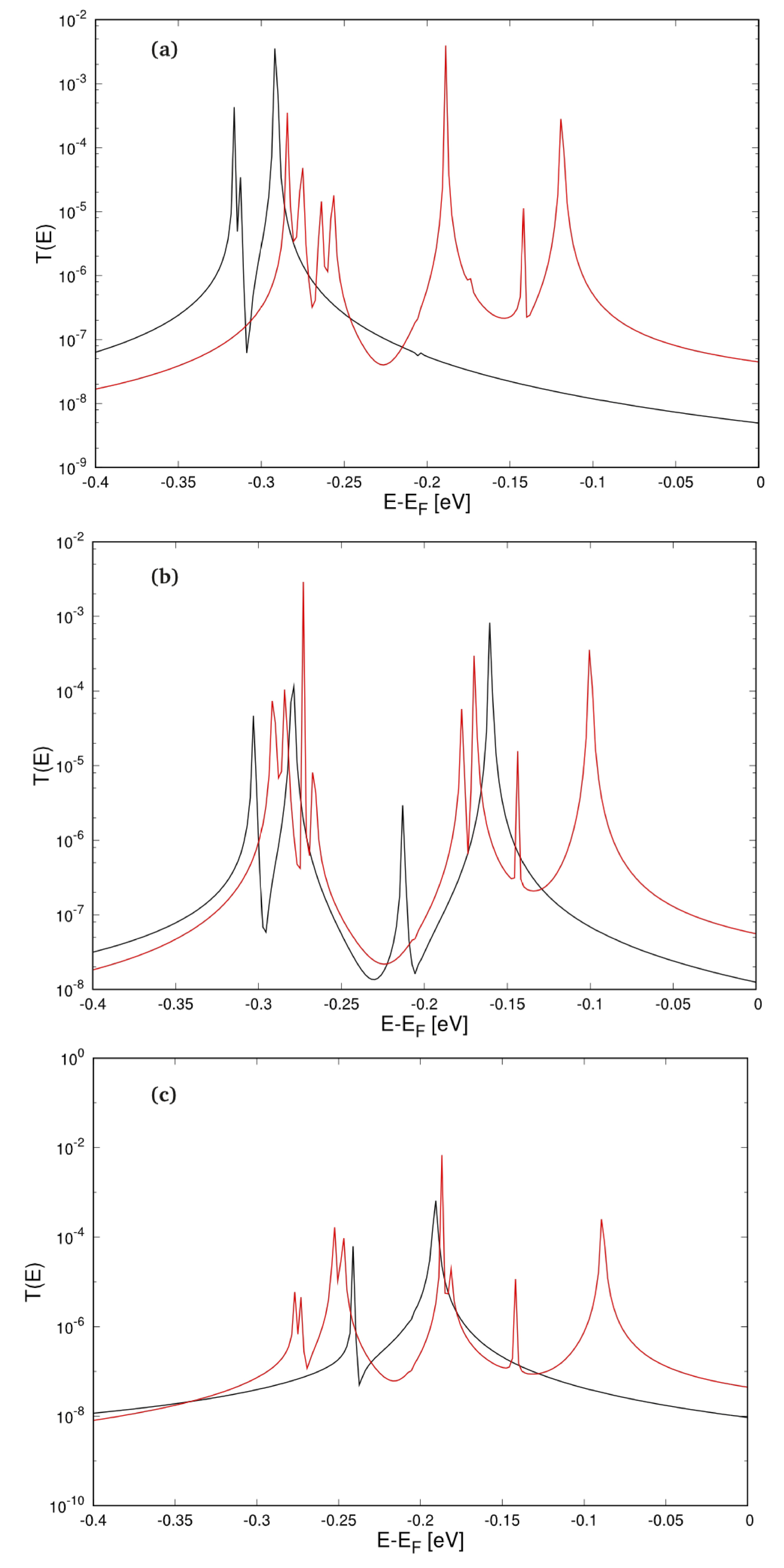}
    \caption{\small Transmission functions for a) Ru/Ru, b) Ru/Os and c) Os/Os in their respective neutral (black solid line) and charged states (red solid line), where for the latter the average of spin-up and spin-down contributions has been used.}
    \label{charged-bigmol-tfs}
    \end{center}
\end{figure}

The redox centers with localized $d$ states on the metal atom in the investigated systems are stabilized by four donor ligands, which suggests a tetragonal ligand field~\cite{J.Bendix 2005}. According to Ligand field theory, for an octahedral field with Jahn Teller distortion, the orbitals $d_{x^2-y^2}$ and $d_{z^2}$ are in higher lying and the orbitals $d_{xz}$, $d_{yz}$ and $d_{xy}$ in lower lying energy levels~\cite{LFT,Georg 2016} (Fig. ~\ref{spin_eigenenergies}a). We find that the peaks in the occupied region close to the Fermi level have contributions mostly from $d$ orbitals with $d_{xz}$ or $d_{yz}$ symmetries (Fig. ~\ref{spin_eigenenergies}b), where the $d_{xy}$ lies further below in energy (Fig. ~\ref{spin_eigenenergies}c). For the systems investigated here the $d$-orbitals $d_{xz}$ and $d_{yz}$ are fully occupied for all systems when in their neutral state, while for the charged state when Ru(II) is ixidized to Ru(III) (Fig. ~\ref{spin_eigenenergies}a), either $d_{xz}$ or $d_{yz}$ is singly occupied according to Hund's rule. This single occupation of a localized state leads to the necessity of spin polarisation in our DFT calculations because spin-up and spin-down orbitals are then not equivalent in energy anymore.

From the respective MO eigenergies within the junction in Fig. ~\ref{spin_eigenenergies}c, which we obtained from a subdiagonalization of the transport Hamiltonian, it can be seen that for the asymmetric system Ru/Os the amount of the shift induced by charging is smaller than for the two symmetric systems Ru/Ru and Os/Os. For Ru/Os also differs from the other two molecules in the energetic sequence of the orbitals, where there are no degeneracies for this case and no changes in sequence when moving from the neutral to the charged system. We also show the difference for the energies of the MOs on the branch closer to Cl and on the other branch with respective $d_{xz}$ and $d_{yz}$ symmetries in Table ~\ref{tab:energydifference}, which confirms the same trends.

\begin{table*}
\begin{center}
\setlength{\tabcolsep}{1.1em}
\caption{Energy differences in eV between the occupied d orbitals close to E$_F$ with symmetries d$_{xz}$ and d$_{yz}$ localized on the two  branches M$_{1}$ and M$_{2}$, where M$_{2}$ is closer to the Cl atom and contains Os for the mixed case Ru/Os. For M$_{1}$ we mark the respective d orbitals as d$_{xz}$ or d$_{yz}$ and for M$_{2}$ as d$^{\prime}_{xz}$ or d$^{\prime}_{yz}$. We use the same notation as in Fig. ~\ref{spin_eigenenergies} for all double branched molecules in their respective neutral and charged states. For all systems we define $\Delta \varepsilon (d_{xz}) = d^{\prime}_{xz} - d_{xz}$ and $\Delta \varepsilon(d_{yz}) = d^{\prime}_{yz} - d_{yz}$.}
\label{tab:energydifference}
    \begin{tabular}{c c c c| c c c}
    \hline
    \hline
    &\multicolumn{3}{ c }{$\Delta \varepsilon(d_{xz})$} &\multicolumn{3}{ c }{$\Delta \varepsilon(d_{yz})$}\\ 
    \hline
      & Ru/Os & Os/Os & Ru/Ru & Ru/Os & Os/Os & Ru/Ru\\
    \hline
     
     neutral & 0.120 & 0.001 &  0.002 & 0.090 & 0.001 & 0.003 \\
    \hline
     charged/spin up & 0.121 & 0.064 & 0.098 & 0.096 & 0.094 & 0.090 \\
    \hline
     charged/spin down & 0.184 & 0.159 & 0.157 & 0.123 & 0.131 & 0.115 \\
    \hline

    \end{tabular}
\end{center}
\end{table*} 

Fig. ~\ref{charged-bigmol-tfs} shows the transmission functions for each branched molecule in this study (Ru/Ru, Ru/Os, Os/Os) in their respective  neutral and charged states, where we conducted spin-polarized NEGF-DFT calculations for all charged  systems. It can be seen that the peak splitting for the two symmetrically built molecules Ru/Ru and Os/Os after charging is pronounced, where MOs on the branch closer to the chlorine atom shifted more with respect to the Fermi level, and the MOs on the other branch almost have not been affected by the chlorine charging effect. For the asymmetrically built molecule Ru/Os the peak splitting caused by charging is less distinct, because of the peak splitting already occuring in the neutral case due to the in-built asymmetry of the molecule, where charging does not seem to increase this splitting much further.

In Table ~\ref{tab:charge and G} we list the conductance of the neutral and charged states for each system as well as the partial charge on each molecule in order to investigate the sources of the asymmetry in MO energies as induced by charging and their effect on the coherent electron transport through the junction. While there is a marked difference between the partial charges on the two branches for the charged versions of all three molecules, the conductance of the charged systems changes only slightly when compared with their neutral counter parts, since the energy shifts of the peaks in the occupied region are relatively small and the conductance is dominated by those rather narrow peaks. Therefore, we conclude that these molecules are not suitable for redox switches, since although one of the two branches can be selectively oxidized this does not result in a significant DQI induced reduction of the conductance.

\begin {table}
\setlength{\tabcolsep}{0.5em}
\caption {Partial charges as obtained from a Bader analysis ~\cite{mulliken} for each of the two branches (metal complex plus all penyl and acetylene spacers but without the pyridyl anchors) of the double branched neutral and charged molecules, where M$_{1}$ and M$_{2}$ denote the branch containing the metal center further away from and closer to the chloride ion, respectively. All values for the charges are given in fractions of electrons. The conductance G for all molecules as defined by $\mathcal{T}(E_{F})$ is given in units of G$_{0}$.} \label{tab:charge and G}
\begin{center}
    \begin{tabular}{c c c c}
    \hline
    \hline
     & M$_{1}$ & M$_{2}$ & G \\ \hline
    \hline
    Ru/Ru (neutral)  & -0.038 & -0.037 & 4.94\texttimes 10$^{-9}$ \\
    \hline
    Ru/Ru (Cl)  & -0.208 & -0.662  & 4.48\texttimes 10$^{-8}$ \\
    \hline
    \hline
    Ru/Os (neutral)  & -0.0345 & -0.0378 & 1.26\texttimes 10$^{-8}$ \\
    \hline
    Ru/Os (Cl) & -0.173 & -0.710  & 5.58 \texttimes 10$^{-8}$ \\
    \hline
    \hline
    Os/Os (neutral)  & -0.036 & -0.036 & 9.19\texttimes 10$^{-9}$ \\
    \hline
    Os/Os (Cl)  & -0.230 & -0.653 & 4.46 \texttimes 10$^{-8}$ \\
    \hline
    \hline
    \end{tabular}
\end{center}
\end{table}
\section{Summary}\label{sec:summary}

In this study we investigated the potential use of branched molecules containing different metal centers in two branches as molecular transistors where the switching would be achieved by a redox process allowing to alternate between an ON and an OFF state, which would differ in their conductance by the occurrence of DQI effects in only one of these two redox states. We did, however, not find a DQI effect in the coherent electron transport through the branched molecules in our study, neither in their neutral nor in their charged states. 

By comparing our results with previously studied ferrocene compounds, we further developed a scheme for the analysis of the structural conditions for the occurence or absence of DQI in branched metal complexes with redox active groups in each branch. We found that the ratio of the through-space coupling t$_{D}$ and the couplings between anchor and bridge states t$_{L/R}$ play a decisive role in this context, which is signficant for chemical design purposes. These parameters, however, are barely altered by the oxidation of one of the two branches. As a consequence, the charging effect on the conductance of these cyclic molecules is not pronounced where only a rather moderate upward shift in energy of the narrow peaks in the transmission curves corresponding to occupied MOs on one branch is found.  

Our findings and the analysis scheme we developed are likely to facilitate the design of DQI-based redox switches and to interpret experimental observations on such complex molecules in the future.

\begin{acknowledgments}
XZ and RS both have been supported by the Austrian Science Fund FWF (project number No. P27272). We are indebted to the Vienna Scientific Cluster VSC, whose computing facilities were used to perform all calculations presented in this paper (project No. 70671). We gratefully acknowledge helpful discussions with Georg Kastlunger.
\end{acknowledgments}

%\newpage
\bibliographystyle{apsrev}

\bibliographystyle{apsrev}

\end{document}